\documentclass{article}
\begin{document}

\noindent {\bf Scaling in the Bombay Stock Exchange Index }
\vskip 1.5 cm

Ashok Razdan\\
Nuclear Research Laboratory\\
Bhabha Atomic Research Centre\\
Trombay, Mumbai-400085\\
\vskip 2.0 cm
\noindent{\bf Abstract:}

In this paper we study BSE Index financial time series for fractal
and multifractal beahiour. We show that Bombay
Stock Exchange (BSE) Index time series
is mono-fractal and can be represented by a fractional Brownian motion.
\vfill\eject
\noindent{\bf Introduction:}

A lot of activity has been witnessed in recent times to study the nature
of finanicial time series. Econophysics [1-4] is an interdisciplinary
field of research in which methods of Physics and Mathematics are applied
to analyse economic systems. Applying Physics to  financial problems
offers fresh look to existing theories of finance.
On the other hand, the availabity of long
term financial data and its behaviour on both at short and  long
time scale offers new interesting challanges to physicsts.
Indeed ,attempts have been made to look for long range and
[3,5] and short range power-law corelation [6].
In this paper we will study scaling behaviour of Bombay Stock Exchange
(BSE) Index of last one year from Ist January 2000 to 31st December 2000
(shown in figure 1).
We will use R/S analysis to look for future trends
and search for multifractal
or Bownian motion features in its time dependence.\\

Fractals have been applied in many fields of sciences like physics,
Biology,Chemical Sciences, Astrophysics and Engineering Sciences.
However ,it is less known that the concept of fractals was born
in the field of economics when  Mandelbrot
in 1963 was investigating the price
changes dynamics of an open market in the year 1963
[ 7]. He found similarity between various
charts of market price changes (of cotton price) with different time resolution.
He came to conclusion that such scale invariance could help to characterise
many complex phenomena seen in physical sciences. Mandelbrot [2] observed
that  price movements follow a family of distributions which have high peaks
and fat tails. Such distributions are known as Stable Paretian  which
have infinite or undefined variance.\\

\noindent{\bf Stock Market Returns: A new paradigm}

Earlier it has been argued that genuinely competitive
stock market, returns ,follow
random walk model and are normally distributed [ 8 ].
Since stock market is a large system and has large
degrees of freedom (or investors),there is a underlyning
assumption that today's change in price is caused only
by today's unexpected "new" information. This means that
today's returns have nothig to do with yesterday's behaviour
and there are no "memory" effects i.e. the returns are
independent. This leads to the argument that the data
of stock prices and returns should follow normal distribution
with stable mean and finite variance. Because of this
arguement, capital market efficiency theories are mainly based
on random walk model or classical Brownian motion concepts.
This approach means that information arrives to an investor
linearly and reaction of an investor to "linear" information 
is instantanous. This is based on assumption that  
yesterday's information has been already folded into yesterday's
price.

However, the actual market data shows that returns are
not normally distributed but have higher peak than theoritically
predicted around the mean
and have fatter tails.
Dow Jones Industrial Index from 1963 to 1993 shows lepokurtic
distrbution [8]. Apart from Dow Jones other stock exchanges
of western countries also show non-normal behaviour [8-11].
The presence of fatter tails indicate
"memory" effects which arise due to non-linear stochastic processes.
Actually the information flow to an investor is
clustered and its arrival  is
irregular rather than continuous and
smooth in nature. This clustered and/or irregular arrival of
"new' information results in periods of low and high volatility [2]
which results in 'leptokurtic' distribution instead of normal.
This brings in a new paradigm in which reaction of investor or
trader to new information is "non-linear".
To investigate the
validity of this new paradigm the concepts of choas theory and fractals
have been used extensively. Models
like ARCH and GARCH [12,13] have been used to include memory
effects but these models
are not popular (mainly from Physics point of view [ 3 ])
because they donot take into account scaling property
of the procress.

\noindent{\bf Brownian Motion and Fractional Brownian motion concepts:}

In economics concept of classical Brownian motion [ 3,9]
have been widely used
to take into account  'memory effects' which get revealed in
the power law behaviour based on random walk model.

Let x(t) be the position of particle
(which is random function of time), than for Brownian motion 
\begin{equation}
x(t)-x(0) =\zeta\|t-t_0|^ h                
\end{equation}
The position x(t) is obtained when $x(t_0)$ is
known and by choosing a random
number $\zeta$ from  a Gaussian distribution.
Here, h=$\frac{1}{2}$ for classical Brownian motion (bm).
In Brownian motion it is not the position of the particle
which is independent of its position at another time but
it is displacement of a given particle at one time which is
independent of its displacement at another time interval.
Brownian motion is 'self-affine'  by nature [ 2 ].
A transformation that  scales time
and distance by different
factors is called affine and behaviour that reproduces itself under
affine transformation is called self-affine [ 2].
Again it was Mandelbrot [ 2 and refrences therein]
who intoduced the concept of fractional Brownian motion (fbm).
Exponent h varies from 0 to 1 in
the above equation for fractional Brownian motion.
For h=$\frac{1}{2}$, the time series is
independent and uncorrelated  but the distribution may not be Gaussian.

\noindent{\bf Hurst Analysis: Search for fractal behaviour}

Hurst invented a new statistical method called, Rescaled range analysis
(R/S analysis) [2,14,15]. He was interested in developing the design of an ideal
reserviour based upon the given record of observed water discharges. Hurst
developed a new exponent called Hurst exponent (H) which can classify time
series into random or non-random . Hurst exponent is also related to fractal
dimension . A measure of a signal "roughness" is also given by Hurst exponent.
The "roughness" of a profile can be defined by observing how signal amplitudes
vary in time (and in space if necessary) ,in particular the correlation
between various amplitude fluctuations. R/S analysis is a method for
distinguishing completely random time series from a correlated time series.
The analysis begins by finding an average over a chosen time period say $\tau$
\begin{equation}
<Z>_{\tau} =\frac{1}{\tau} \sum_{t=1}^{\tau} Z(t)
\end{equation}
Let X(t) be the accumlated deparature of the influx Z(t) from the mean
$<z>_{\tau}$.
\begin{equation}
X(t,{\tau})=\sum_{u=1}^{t} (z(u)-<z>_{\tau})
\end{equation}
The difference between the maximum and the minimum accumlated influx X
is the range R which is given as
\begin{equation}
R({\tau})=max X(t,{\tau})-min X(t,{\tau})
\end{equation}
Hurst used the dimensionless ratio $\frac{R}{S}$, where S is the
standard deviation which is given as
\begin{equation}
S=(\frac{1}{\tau} \sum_{t=1}^{\tau} (Z(t)-<z>_{\tau})^2)^{\frac{1}{2}}
\end{equation}
Hurst found that the observed rescaled range, $\frac{R}{S}$ ratio
for a time series is given as
\begin{equation}
\frac{R}{S}= (\frac{\tau}{2})^H
\end{equation}
where H is the Hurst exponent.
Using the above equation we can find exponent H.
If H is between 0.5 and 1 , the trend is persistent  which indicates
long memory effects . This also means that the increasing trend in
the  past implies increasing trend in the future also or decreasing
trend in the past implies decreasing trend in the future also.
In contrast to this ,if H is between 0 and 0.5 than an increasing
trend in the past implies an decreasing trend in future and decreasing
trend in past implies increasing trend in future. It is important to
note that  persistent stochastic processes have little noise whereas
anti-persistent  processes show, presence of high frequency noise.

The relationship
between fractal dimensions $D_f$ and hurst exponent H can expressed
as [11]
\begin{equation}
D_f= 2-H
\end{equation}

So by finding Hurst exponent of a financial time series,we can find
out the fractal dimension of the time series.When $D_f$=1.5 ,there is
normal scaling. When $D_f$ is between 1.5 and 2,
time series is anti-persistent
and when $D-f$ is between 1 and 1.5 the time series is 
persistent. For $D_f$=1,time series is smooth curve and purely
deterministic in nature and for $D_f$=1.5 time series is purely random.
Long term correlations of indexes in developed and
emerging markets have been studied by using Hurst analysis and
detrended fluctuation analysis (DFA) as investigating tools [24].
However, it has been argued by Vandewalle and Ausloos [25]
that DFA analysis is better than Hurst scaling for
short term time series.

\noindent {\bf Global Hurst Exponent: Search for multifractal nature}

In general box counting method is being used for studying
multifractal features [15,16,17,18] but this method is mainly
prefered for problems of spatial nature. However,
to study multifractal nature of financial time
series, alternative methods  have been suggested [19].
This technique involves calculating qth order height-height correlation
function or qth order structure function [19,20] of a normalized time
series y($t_i$)
\begin{equation}
c_q(\tau)=<|y(t_{i+r})-y(t_i)|^{q}>
\end{equation}
where only non-zero terms are considered in the average,taken over
all pairs ($t_{i+r},t_i$) such that

\begin{equation}
\tau=|t_{i+r} -t_i|
\end{equation}
and
\begin{equation}
c_q(\tau) \sim {\tau}^ {\eta(q)}
\end{equation}

where q $\ge$ 0 is the order of moment  and $\eta(q)$ is the scale
invarient structure function exponent. For q=1, H=$\eta(1)$ is the
"Hurst" exponent characterizing the scaling non connservation of
mean. For q=2, we obtain $\eta(2)$=$\beta$-1,where $\beta$ is the
slope of the fourrier power spectrum. In general $\eta(q)$ is given
by
\begin{equation}
\eta(q)= qH -{\frac{C_1}{\alpha-1}} (q_{\alpha}-q)
\end{equation}

where $C_{1}\le$ d is an intermittency parameter, d is the dimension of
space( here d=1) and $\alpha$ varies between 0 and 2. $\alpha$ is the
Levy index.
A multifractal process is characterized by a non-linear behaviour
of $\eta(q)$ [21] because  of multiplicative cascades  where as
those processes which are additive in nature, $\eta(q)$ is linear
or bi-linear. For Brownian motion (bm), $\eta(q)$=$\frac{q}{2}$ and
for fractional Brownian motion (fbm), $\eta(q)$=qH [22,23].Thus for a purely
bm of fbm ,$\eta(q)$ is linear, whereas for multifractal nature $\eta(q)$
is non-linear.\\

\noindent{\bf Data Analysis and Results:}

Following the above discussion , we analyse BSE index data of year 2000
from Ist january to 31st December which gives us 245 data points. Data
from time financial series are one dimensional and more simple to anlyze
than spatial one. For a financial time series there are no holidays or
weekends. In order to do R/S analysis data has been divided into 20,40,
.... 220,240 parts . The next step is to calculate R/S statistics and
plot it against the corresponding sample length on double lograithmic
plot as shown in the figure 2.
The Gaussian asymptotic behaviour  of R/S which represents independent
random process with finite variances is given in line marked 'b' in figure 2
and can be written as
\begin{equation}
\frac{R}{S}=(\frac{\pi \tau}{2})^{\frac{1}{2}}
\end{equation}
It is clear from this figure that data fits
to the Hurst exponent H = 0.915 .
This value has been obtained by fitting  $\frac{R}{S}$=$(a\tau)^{H}$
to the observed R/S value . The parameter a= 0.61 for the fit
given in the line marked a in figure 2.
This value of H is an example of persistent
behaviour. Fractal dimension for BSE index time series is $D_f$=1.085.
Large values of H , have been obtained for many naturally
occuring phenomena like monthly sunspot activity.
In figure 3 we have plotted $\eta(q)$ =qH(q),  with q which
has obtained by calculating structure function. 
We find that $\eta(q)$ has a linear relationaship with q which
as discussed above shows that BSE index does not have multifractal
nature but it represents a fractional Brownian motion.\\

\noindent{\bf Conclusions:}

By using $\frac{R}{S}$ analysis ,we are able to find fractal dimension
of of BSE Index. We also observed that trend in BSE index is Perisitent
for the Year 2000.  This means if market is not doing well in the year
2000, persistent trend will continue i.e. in future, market will 
continue to give low returns. If we see behaviour of BSE from January
2001 onwards, market perfomance has not been good.
The linear behaviour of qH(q) values of BSE Index
with q shows that signal is mono-fractal and data follows simple
scaling values for these values of q. But it important to note that
we have used data of only one year comprising 240 data points. It will be
intresting to look for multifractal features in Short term ( single day
data, but intraday behaviour)
or long term (may be decade or more) data of BSE Index.In other stock
exchanges such studies have shown multifractal features [24]. It is
important to make such studies because market returns have been 
correlated to [24] multifractal features in the Index data,more so
when functioning of BSE will be more transparent from July 2001 onwards,
with the introduction of new market mechanisms like "OPITIONS" and
"FUTURES".\\
\noindent{\bf Acknowledgement:} I am thankful to Dr S.K.Chiragi for
his inputs, which helped me in the improvements of this paper.

\vfill\eject
\noindent{\bf Figure Captions:}

Figure1: It shows BSE index for the Year 2000 from Ist January 2000 to
31 st December 2000 [26]. \\

Figure2: It shows Log( R/S) with log($\tau$). Curve 'a' corresponds
to actual data and curve 'b' to asymptotoc theoritical expectation.\\

Figure3: $\eta(q)$ =q H(q), has been plotted against q for BSE index.
It shows linear behaviour indicating fractional Brownian motion.\\

\vfill\eject
\noindent{\bf References:}
\begin{enumerate}
\item  J.M. Pimbly , Phys. Today 50(1997)42. 
\item  B.B. Mandelbrot, Fractals and Scaling in Finance 
       (Springer,New York ,1997 edition)
\item  N.VandeWalle, M. Ausloos in Econophysics -an Emerging Sciences
       Edited by J.Kertesz,I.Kondor (Kluwer,Dordrecht 1999 Edition)
\item  R.N.Mantenga, H.E.Stanely , Nature 376(1995)46;ibid.,
       Physica A 239(1997)255;ibid.,Physica A 254(1998)77 
\item  B.Chopard, R.Chatangny in Scale Invariance and Beyond, Edited by
       B.Dubrulle, F.Graner, D.Sornette (EDP Sciences 1997)
\item  K.Ivanova, M.Ausloos, Physica A 265(1999)279
\item  B.B. Mandelbrot, J. of Business (Chicago)36 (1963)394
\item  E.F.Femma, J.Finance (1970) 383-417
\item  R.N.Mantegne, H.E. Stanely , An Introduction to Econophysics
       Cambridge University Press ,2000.
\item  E.E.Peters, Chaos and Order in Capital Markets, Wiley,New York,1991
\item  E.E.Peters,  Fractal Market Analysis,Wiley,New York,1994
\item  J.D.Hamilton, Time Series Analysis,Princeton University Press,NJ,1994
\item  R.Engle, Econometrica 50(1982)987
\item  H.E.Hurst, Long Term Storage of Reseviours, Trans. Amer. Soc.
       civil Eng. 116(1950)770-808
\item  R. Weron,B.Przybylowicz, Physica A 283(2000)462
\item  A.B.Chhabra,C.Meneveau, R.V.Jensen and K.R.Sreenivasan,
       Physical Review A 40(1989)5284
\item  A.Haungs,A.Razdan,C.L.Bhat,R.C.Rannot,H.Rebel, Astroparticle
       Physics 12(1999)145
\item  A.Razdan, A.Haungs,H.Rebel and C.L.Bhat, Astroparticle Physics(submitted)
\item  A.L.Barabasi, T.Vicsek, Phys.Rev. A 44(1991)2730
\item  F.Schmitt,D.Schertzer,S.Lovejoy , Int. J.Theor. Appl. Fin.,3(2000)361
\item  G.Parisi  and U.Frisch in Turbulance and Predicibility in Geophysical
       fluid dynamics and Climate dynamics,eds M. Ghil et. al.(North-Holland)
       1955,pp84-92
\item  B.J.West, The Lure of Modern Science:Fractal Thinking (World Scientific
       1985)
\item  M.V.Berry ,Z.V.Lewis,Proc. R. Soc. London A370(1980)459
\item  M.Beben and A.Orlowski, Euro. Phys. J. B 20(2001)527
\item  N.Vandewalle and M.Ausloos, Physica A 246(1997)454
\item  www.bseindia.com
\end{enumerate}

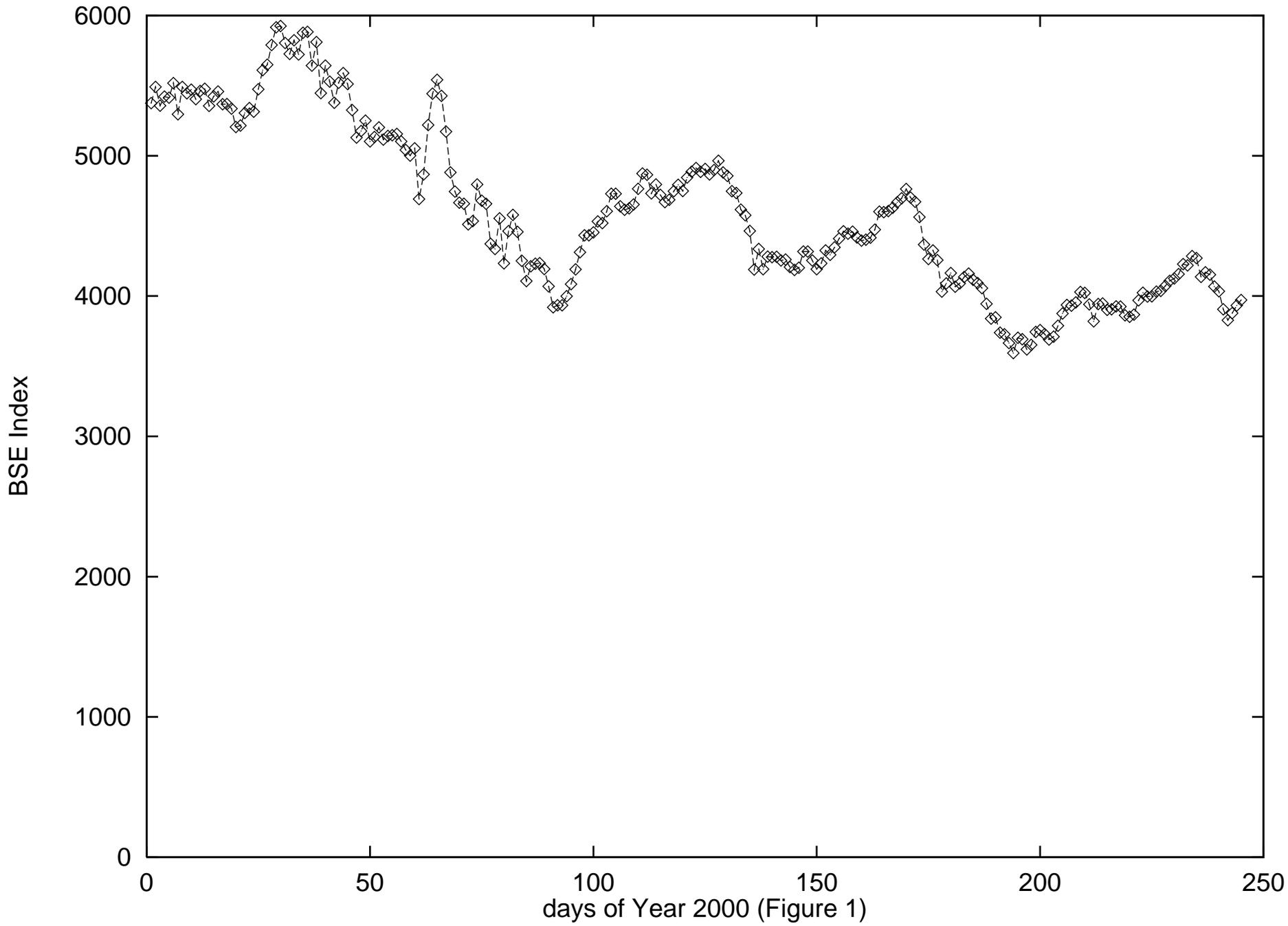
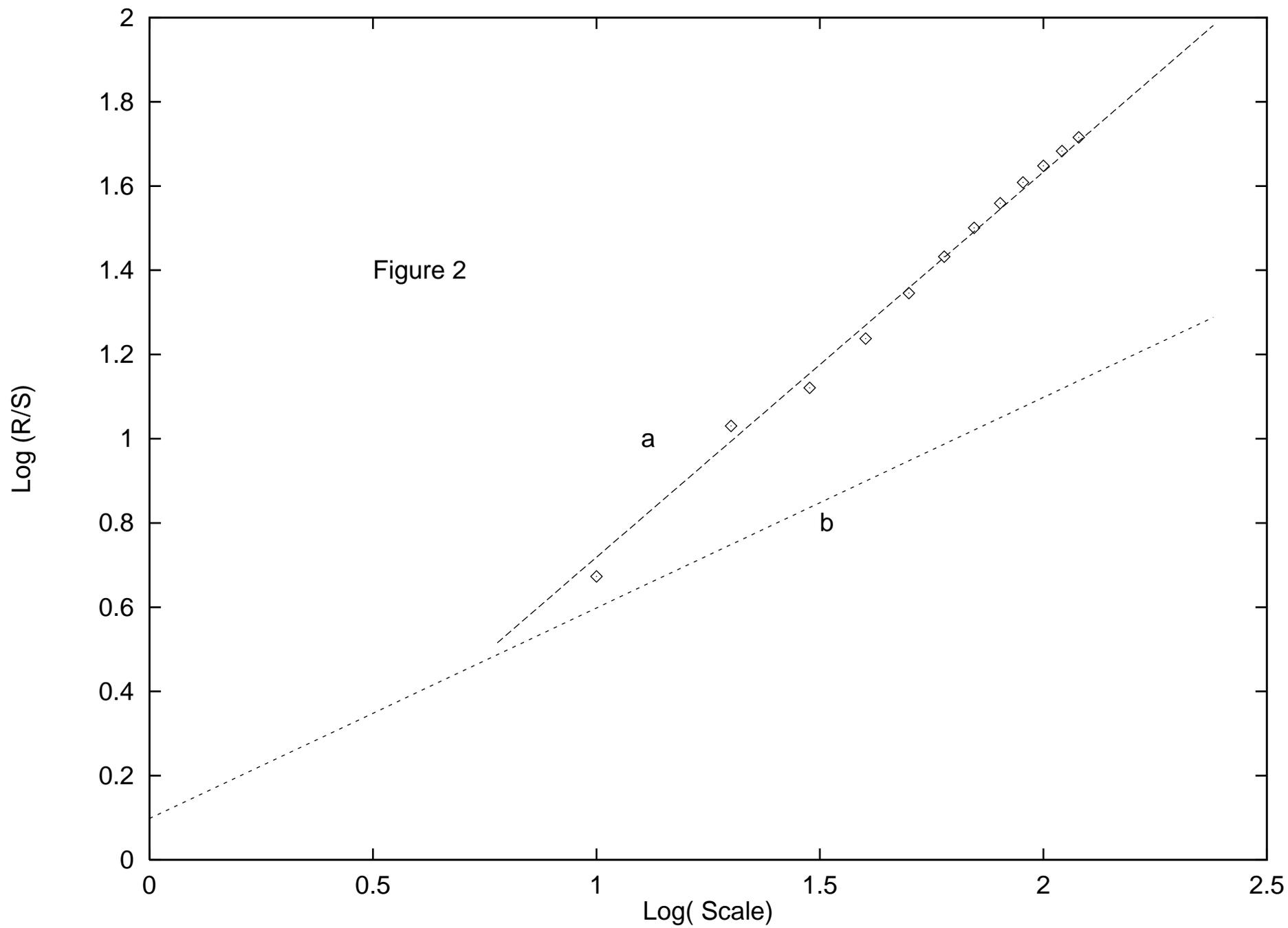
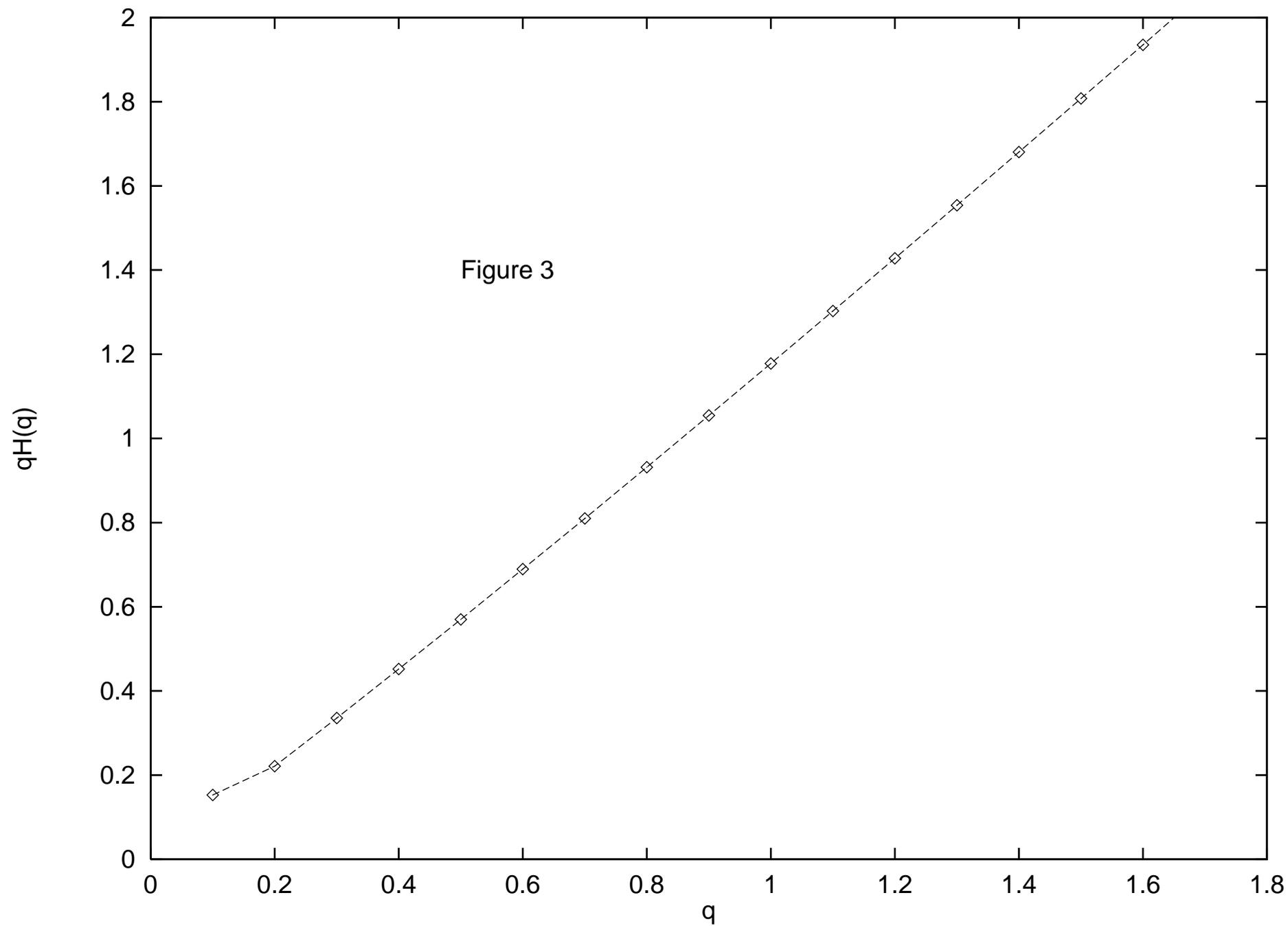
\end{document}